\documentclass[superscriptaddress,
twocolumn,showpacs,preprintnumbers,amsmath,amssymb]{revtex4}
\usepackage{graphicx}% Include Figure files
\usepackage{dcolumn}% Align table columns on decimal point
\usepackage{bm}% bold math
\usepackage[usenames]{color} 
\usepackage{latexsym}
\begin{document}
\title{RNA polymerase motor on DNA track:\\
effects of interactions, external force and torque}
\author{Tripti Tripathi}
\affiliation{Physics Department, Indian Institute of Technology, Kanpur 208016, India}
\author{Prasanjit Prakash}
\affiliation{Physics Department, Indian Institute of Technology, Kanpur 208016, India}
\author{Debashish Chowdhury{\footnote{Corresponding author; E-mail: debch@iitk.ac.in}}}
\affiliation{Physics Department, Indian Institute of Technology, Kanpur 208016, India}
%\email{debch@iitk.ac.in}
\date{\today}%
%%%%%%%%%%%%%%%%%%%%%%%%%%%%%%%%%%%%%%%%%%%%%%%%%%%%%%%%%%%%%%%%%%%%%%%%%%%%%%%%%%%%%%%%%
\begin{abstract} 
RNA polymerase (RNAP) is a mobile molecular workshop that polymerizes 
a RNA molecule by adding monomeric subunits one by one, while moving step 
by step on the DNA template itself. Here we develop a theoretical model by 
incorporating the steric interactions of the RNAPs and their mechanochemical 
cycles which explicitly captures the cyclical shape changes of each motor. 
Using this model, we explain not only the dependence of the average velocity 
of a RNAP on the externally applied load force, but also predict a {\it 
nonmotonic} variation of the average velocity on external torque. We also 
show the effect of steric interactions of the motors on the total rate of 
RNA synthesis. In principle, our predictions can be tested by carrying out 
{\it in-vitro} experiments which we suggest here.
\end{abstract}
%%%%%%%%%%%%%%%%%%%%%%%%%%%%%%%%%%%%%%%%%%%%%%%%%%%%%%%%%%%%%%%%%%%%%%%
\maketitle
%%%%%%%%%%%%%%%%%%%%%%%%%%%%%%%%%%%%%%%%%%%%%%%%%%%%%%%%%%%%%%%%%%%%%%%
\section{Introduction} 
%%%%%%%%%%%%%%%%%%%%%%%%%%%%%%%%%%%%%%%%%%%%%%%%%%%%%%%%%%%%%%%%%%%%%%%%%%%%%

Molecular motors \cite{schliwa} in living cells are either proteins or 
macromolecular complexes made of proteins and ribonucleic acids (RNAs). 
Like their macroscopic counterparts, these motors perform mechanical work, 
while translocating on a filamentous track, by converting input energy 
which is often supplied as chemical energy \cite{schliwa,howard,fisher}. 
In this paper we study a special class of motors, called RNA polymerase 
(RNAP) which play crucial roles in gene expression \cite{alberts}.

Transcription of a gene encoded in the sequence of nucleotides in a 
specific segment of a DNA is carried out by RNAP motors which treat the 
DNA as a template \cite{borokhov08,korzheva03}. An RNAP is more like a 
mobile workshop that performs three functions simultaneously: (i) it 
decodes the genetic message encoded in the template DNA and selects the 
appropriate nucleotide, the monomeric subunit of RNA, as dictated by the 
template, (ii) it catalyzes the addition of the monomeric subunit thus 
selected to the growing RNA molecule, (iii) it steps forward by one 
nucleotide on its template without completely destabilizing the ternary 
complex consisting of the polymerase, the template DNA and the product 
RNA. The free energy released by the polymerization of the RNA molecule 
serves as the input energy for the driving the mechanical movements of 
the RNAP. Therefore, these enzymes are also regarded as molecular motors 
\cite{gelles98}. 

During the transcription of a gene, the collective movement of the RNAPs 
on the same DNA track is often referred to as RNAP traffic because of its 
superficial similarity with vehicular traffic \cite{polrev,css}. 
The beginning and the end of the specific sequence corresponding to a 
gene are the analogs of the on-ramp and off-ramp of vehicular traffic 
on highways. The average number of RNAPs, which complete the synthesis 
of a RNA molecule per unit time interval can be identified as the 
{\it flux} in RNAP traffic. Note that flux is the product of the number 
density and average velocity of the motors. Thus, flux in RNAP traffic 
is identical to the average rate of synthesis of the corresponding RNA. 
Using the terminology of traffic science \cite{css}, we'll call the 
relation between the flux and the number density of the motors as the 
{\it fundamental diagram}. The fundamental diagram is an important 
quantitative characteristic of traffic flow. 

The dependence of the velocity of the motor on an externally imposed 
load (opposing) force is called the force-velocity relation which is 
one of the most important characteristics of a molecular motor. The 
force-velocity relation for RNAP motors have been measured by carrying 
out single molecule experiments \cite{bai06,herbert08}. However, to our 
knowledge, the response of an RNAP motor to an externally applied 
torque has not been investigated so far. The effects of steric 
interactions of the RNAP motors on their dynamics has been studied only in 
a few experiments \cite{bremer95,nudler03a,nudler03b,crampton06,sneppen05}; 
but, none of these addressed the question of the nature of the overall 
spatio-temporal organization of the RNAP motors in RNAP traffic. 

The traffic-like collective dynamics of cytoskeletal molecular motors 
\cite{lipo,frey,santen,popkov1,nosc,greulich}
and that of ribosomes on mRNA tracks 
\cite{macdonald68,macdonald69,lakatos03,shaw03,shaw04a,shaw04b,chou03,chou04,schon04,schon05,dong,basuchow} 
have been investigated theoretically in the physics literature. However, 
so far, RNAP traffic has received far less attention 
\cite{liverpool,ttdc,klumpp08} 
In this paper we develop a model that captures not only the steric 
interactions between the RNAPs, but also separately the {\it 
biochemical reactions} catalyzed by an RNAP and the {\it cyclic shape  
changes} it undergoes during each mechano-chemical 
cycle. This model may be regarded as a ``unified'' description in the 
sense that the same model describes the single RNAP properties (e.g., 
the force-velocity and torque-velocity relations) as well as the 
collective spatio-temporal organization (and rate of RNA synthesis).

%%%%%%%%%%%%%%%%%%%%%%%%%%%%%%%%%%%%%%%%%%%%%%%%%%%%%%%%%%%%%%%%%%%%%%%
\section{Basic mechano-chemistry and the model}
%%%%%%%%%%%%%%%%%%%%%%%%%%%%%%%%%%%%%%%%%%%%%%%%%%%%%%%%%%%%%%%%%%%%%%%

The main stages in the polymerization of polynucleotides by the polymerase
machines are common:\\
(a) {\it initiation}: Once the polymerase encounters a specific sequence
    on the template that acts as a chemically coded start signal, it
    initiates the synthesis of the product. The RNAP, together with the 
    DNA bubble and the growing RNA transcript, forms a ``transcription 
    elongation complex'' (TEC). This stage is completed when the nascent 
    product RNA becomes long enough to stabilize the TEC against 
    dissociation from the template. \\
(b) {\it elongation}: During this stage, the nascent product gets
    elongated by the addition of nucleotides; during elongation 
    \cite{uptain}, each successful addition of a nucleotide to the 
    elongating mRNA leads to a forward stepping of the RNAP. \\
(c) {\it termination}: Normally, the process of synthesis is terminated,
    and the newly polymerized full length product molecule is released,
    when the polymerase encounters the {\it terminator} (or, stop)
    sequence on the template. \\
In this paper we are interested mainly in the elongation of the 
mRNA transcripts.

%%%%%%%%%%%%%%%%%%%%%%%%%%%%%%%%%%%%%%%%%%%%%%%%%%%%%%%%%%%%%%%%%%%%%%%%%%%
\begin{figure}[h]
\begin{center}
\includegraphics[width=0.75\columnwidth]{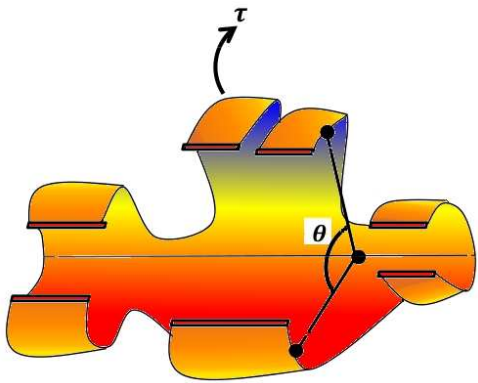} 
\end{center}
\caption{(Color online) A schematic depiction of the essential architectural 
feature of a RNAP that we capture in the model proposed in this paper. The 
RNAP can switch between ``open'' and ``closed'' shapes cyclically in 
each mechano-chemical cycle of its operation. The extent of ``opening'' 
is measured by the angle $\theta$ which can be manipulated also by 
externally applied torque $\tau$.
}
\label{fig-architec}
\end{figure}
%%%%%%%%%%%%%%%%%%%%%%%%%%%%%%%%%%%%%%%%%%%%%%%%%%%%%%%%%%%%%%%%%%%%%%%%%%%

%%%%%%%%%%%%%%%%%%%%%%%%%%%%%%%%%%%%%%%%%%%%%%%%%%%%%%%%%%%%%%%%%%%%%%%%%%%
\begin{figure}[h]
\begin{center}
\includegraphics[angle=-90,width=0.75\columnwidth]{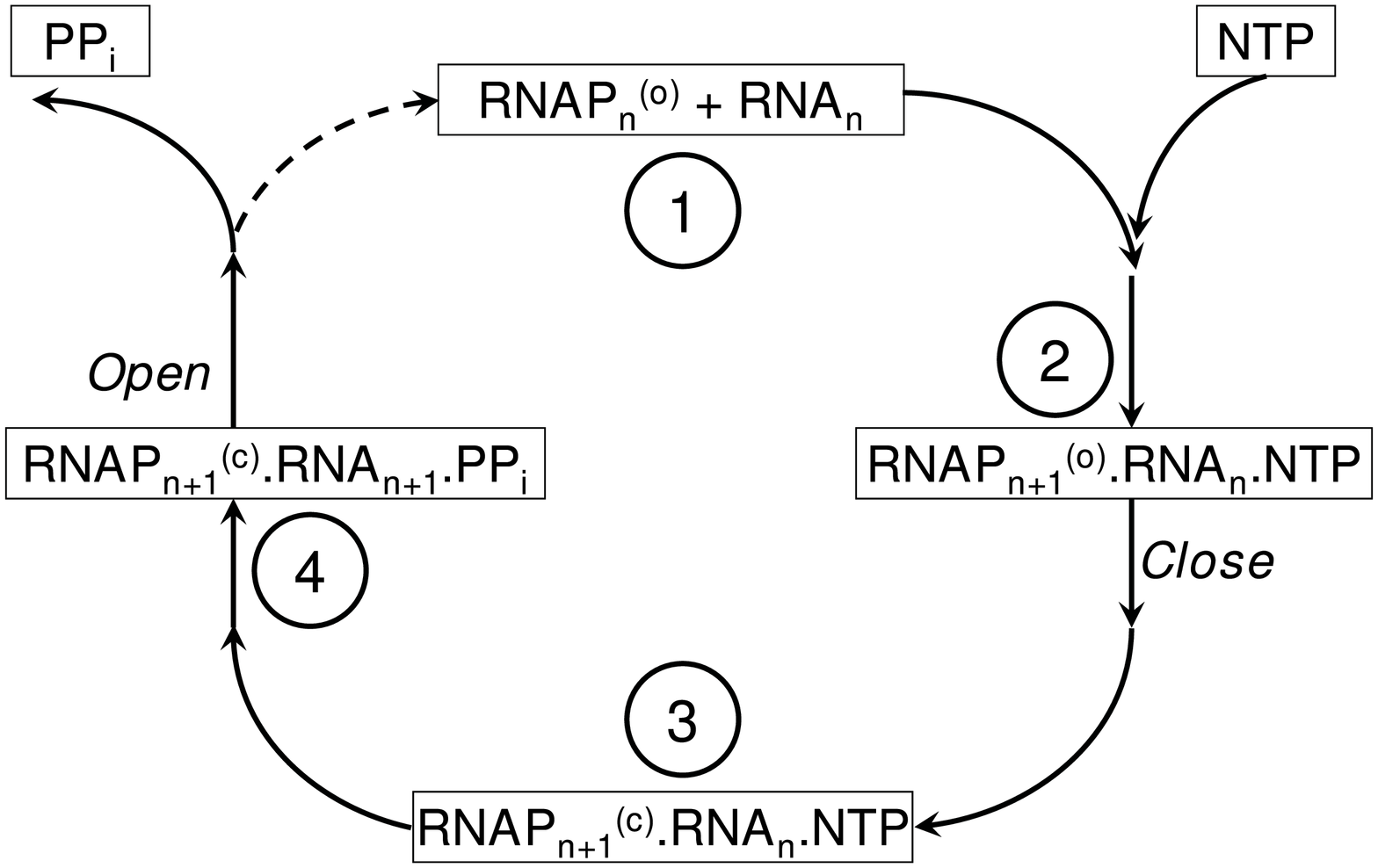} 
(a)
\includegraphics[angle=-90,width=0.75\columnwidth]{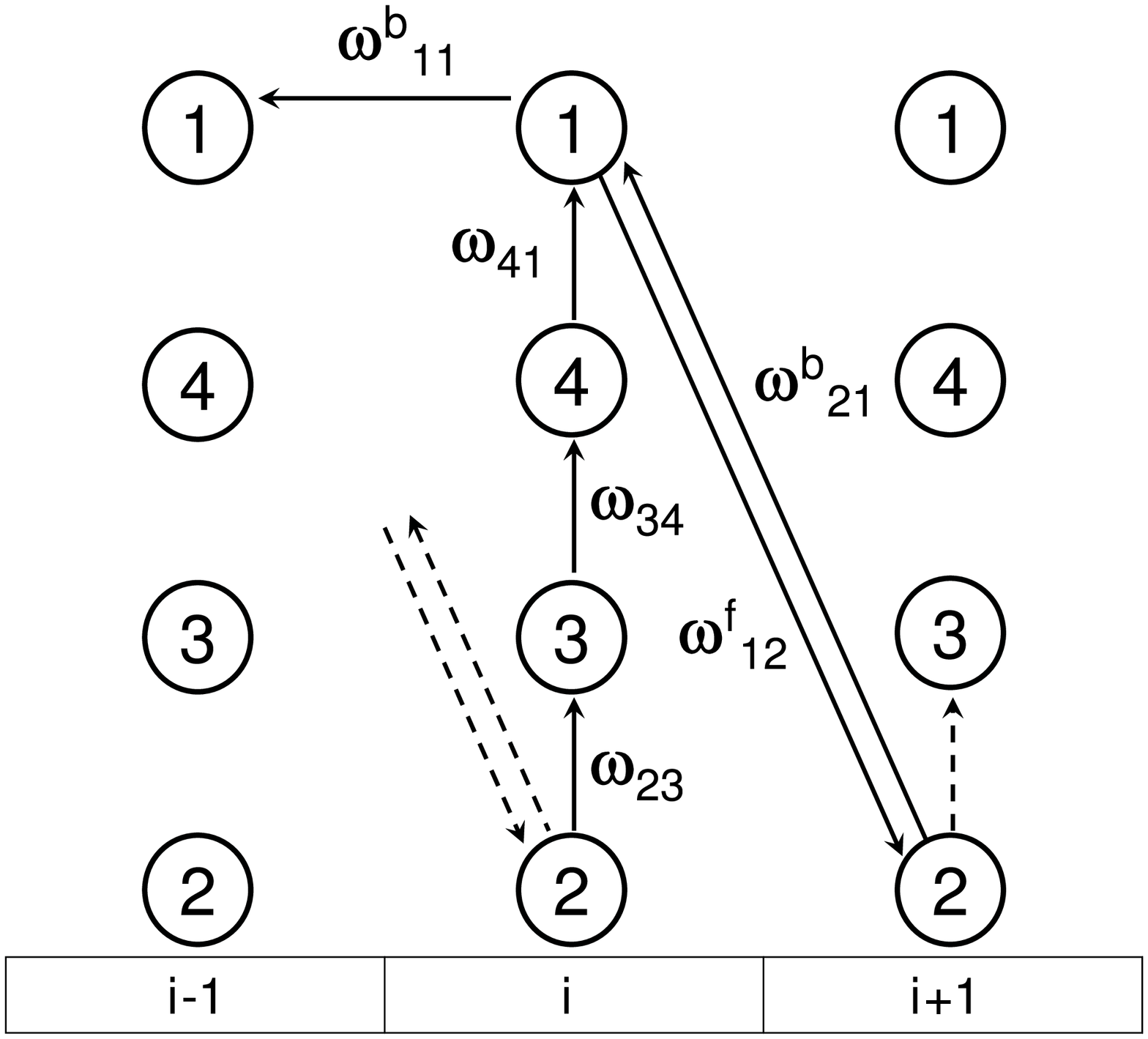}
(b)
\end{center}
\caption{ A schematic representation of the mechano-chemical cycle of
each RNAP in our model in the elongation stage is shown in (a) (see 
the text for explanation). The 
labels $...,i-1, i, i+1,...$ in (b) denote the nucleotides on the 
DNA template. Thus, the horizontal direction in (b) denote positions 
of the RNAP while the vertical direction, at a given position, describe 
the corresponding ``chemical'' states. 
}
\label{fig-model}
\end{figure}
%%%%%%%%%%%%%%%%%%%%%%%%%%%%%%%%%%%%%%%%%%%%%%%%%%%%%%%%%%%%%%%%%%%%%%%%%%%

One of the key experimental observations is that an RNAP cycles between 
a ``closed'' and an ``open'' shape during each mechano-chemical cycle 
(see fig.\ref{fig-architec} and fig.\ref{fig-model}a) 
\cite{temiakov04,yin04,guo06,borukhov08}. 
Recognition of the correct incoming nucleotide (dictated by the template) 
and its binding with the catalytic site on the RNAP leads to its closing. 
It remains closed as long as the hydrolysis of the NTP takes place. The 
nascent RNA is elongated by one monomer on completion of this reaction 
which also produces pyrophosphate. Then, the shape of the RNAP switches 
back to its open shape from the closed one thereby facilitating release 
of the pyrophosphate. In the open shape, the RNAP is weakly bound to its 
DNA track and, in principle, can execute Brownian motion. 

To our knowledge, none of the models of RNAP traffic reported earlier, 
explicitly capture these cyclic change of shape of the polymerase motor.
Liverpool et al.\cite{liverpool} as well as Klumpp and Hwa \cite{klumpp08} 
treat each RNAP as a rigid rod and describe its translocation from one 
nucleotide to the next on its template in terms of a single parameter. 
Thus, both these models capture the effects of the mechano-chemistry of 
indivisual RNAPs by an effective ``hopping'' rate contstant.
In an attempt to capture the most essential aspects of the mechano-chemical 
cycle of individual RNAP motors, we have assigned two possible ``internal''  
states to an RNAP at each spatial position on its track \cite{ttdc}. 
However, each of these two states is, in reality, not a single bio-chemical 
state; a sequence of bio-chemical states which interconvert sufficiently 
rapidly are collectively represented by one ``internal'' state whereas 
the transition from one internal state to the other is the slowest of all 
the transitions and, therefore, rate-limiting. This minimal model neither 
distinguishes between purely chemical transitions pure shape changes, nor 
does it assign distinct internal states to the ``open'' and ``closed'' 
shapes of the RNAP motor. Therefore, in this paper, we extend our 
earlier 2-state model to a 4-state model.

We shall use the terms RNAP and TEC interchangeably. We extend the 
2-state model of ref.\cite{ttdc} by assigning four possible distinct 
``internal'' state to the RNAP  to capture its transitions between open 
and closed shapes. The composition of these four states and the spatial 
position of the RNAP as well as the transitions between these states are 
shown in fig.\ref{fig-model}(b). 

The subscript $n$ of the symbol ``RNAP'' denotes the position of the 
RNAP motor on its track, measured from a nucleotide that marks the  
starting point of transcription by the RNAP. The two superscripts 
$(o)$ and $(c)$ of ``RNAP'' refer to its ``open'' and ``closed'' shapes, 
restectively. The four distinct states in each cycle are labelled by 
the integers 1,2,3,4. Let us begin with the state 1, where the RNAP is 
open and is located at $n$ so that the length of the corresponding 
elongating mRNA transcript is also $n$. Arrival of the correct NTP 
subunit causes the transition to the state 2 where the open RNAP, now 
located at the new position $n+1$,  forms a complex with the mRNA 
transcript of length $n$ and the newly arrived NTP subunit. Next, 
the shape of the RNAP changes from open to closed form and the new 
state of the system is labelled by the integer index 3. Then, the 
closed RNAP catalyzes the formation of a new covalent bond between 
the mRNA transcript and the newly arrived NTP subunit which leads to 
the elongation of the transcript from $n$ to $n+1$ and the corresponding 
transition from state 3 to state 4. Finally, the RNAP opens again and 
releases the pyrophosphate ($PP_i$); this transition takes the open 
RNAP to the state 1 at the position $n+1$. The 4-state model reduces 
to our earlier 2-state model \cite{ttdc} in the limit 
$\omega_{23}\rightarrow\infty$,$\,\omega_{34}\rightarrow\infty$\\\\

%%%%%%%%%%%%%%%%%%%%%%%%%%%%%%%%%%%%%%%%%%%%%%%%%%%%%%%%%%%%%%%%%%%%%%%%%%%%%%
\section{RNAP TRAFFIC UNDER PERIODIC BOUNDARY CONDITION}
\label{sec-pbc}
%%%%%%%%%%%%%%%%%%%%%%%%%%%%%%%%%%%%%%%%%%%%%%%%%%%%%%%%%%%%%%%%%%%%%%%%%%%%%%

In our model a one dimensional lattice of $L$ sites represents the DNA 
template. Each site of these lattice corresponds to a single base pair.
The total number of RNAP motors which move simultaneously on this 
template (and transcribe the same gene sequentially) is $N$; the linear 
size of each RNAP is $r$ in the units of base pairs. Therefore, the 
{\it number density} of the RNAP motors is $\rho = N/L$ whereas the 
corresponding {\it coverage density} is $\rho_{cov} = N r/L$. 
The instantaneous spatial position of a RNAP is denoted by the leftmost 
site of the lattice it covers at that instant of time; however, it also 
``covers'' all the adjacent $r-1$ on the right side of of the site which 
denotes its position. In order to capture the steric interaction (mutual 
exclusion) of the RNAPs, no site of the lattice is allowed to be be 
covered simultaneously by more than one RNAP. Thus, this model of RNAP 
traffic may be regarded as a totally asymmetric simple exclusion process 
(TASEP) \cite{sz,schuetz} for hard rods each of which can exist in one of 
its four possible ``internal'' states at any arbitrary position on the 
lattice.

Let $P_{\mu}(i,t)$ denote the probability that there is a RNAP at the
spatial position $i$ and in the chemical state $\mu$ at time $t$.
At any arbitrary site $n$ and time $t$, these probabilities must satisfy 
the normalization condition
\begin{eqnarray}
P(n,t) = \sum_{\mu=1}^{4} P_{\mu}(n,t) = \frac{N}{L} = \rho 
\label{eq-normal1}
\end{eqnarray}
Under the mean-field approximation, the master equations for 
$P_{\mu}(i,t)$ are as follows:
%%%%%%%%%%%%%%%%%%%%%%%%%%%%%%%%%%%%%%%%%%%%%%%%%%%%%%%%%%%%%%%%%%%%%%%
\begin{widetext}
\begin{eqnarray}
\frac{dP_{1}(n,t)}{dt}&=&P_{1}(n+1,t)~\omega_{11}^{b}\,Q(n+1-r|\underline{n+1})+P_{2}(n+1,t)~\omega_{21}^{b}~Q(n+1-r|\underline{n+1})\nonumber\\
&+&P_{4}(n,t)~\omega_{41}-P_{1}(n,t)\left[~\omega_{11}^{b}\,Q(n-r|\underline{n})+~\omega_{12}^{f}~Q(\underline{n}|n+r)\right]
\label{eq-p1}
\end{eqnarray}
\begin{eqnarray}
\frac{dP_{2}(n,t)}{dt}&=&P_{1}(n-1,t)~\omega_{12}^{f}~Q(\underline{n-1}|n-1+r)~-~P_{2}(n,t)\left[ ~\omega_{21}^{b}~Q(n-r|\underline{n})+~\omega_{23}\right]
\label{eq-p2}
\end{eqnarray}
\begin{eqnarray}
\frac{dP_{3}(n,t)}{dt}&=&P_{2}(n,t)~\omega_{23}-P_{3}(n,t)~\omega_{34}
\label{eq-p3}
\end{eqnarray}
\begin{eqnarray}
\frac{dP_{4}(n,t)}{dt}&=&P_{3}(n,t)~\omega_{34}-P_{4}~\omega_{41}
\label{eq-p4}
\end{eqnarray}
\end{widetext}
where $Q(\underbar{i}|j)$ is the conditional probability that, given an 
RNAP in site i, site j is empty.
We calculate $Q(\underline{n}|n+r)$ following the same steps as we did 
in ref.\cite{ttdc}, getting 
\begin{equation}
Q(\underline{n}|n+r)=Q(n|\underline{n+r})=\frac{1-\rho\,r}{1+\rho-\rho\,r}
\end{equation}

In the steady state, all the $P_{\mu}(n,t)$'s  become independent of time. 
Moreover, because of the periodic boundary conditions, all the sites are 
equivalent so that the steady-state probabilities $P_{\mu}(n)$'s are also 
independent of site index $n$. 
Solving Eqn.(\ref{eq-p1},\ref{eq-p2},\ref{eq-p3},\ref{eq-p4}), together 
with normalization condition (\ref{eq-normal1}), in the steady state 
under periodic boundary conditions, we get
\begin{eqnarray}
P_{1}&=&\rho\frac{\omega_{23}~\omega_{34}~\omega_{41}+\omega_{34}~\omega_{41}~\omega_{21}^{b}Q}{K}\\
P_{2}&=&\rho\frac{\omega_{34}~\omega_{41}~\omega_{12}^{f}Q}{K}\\
P_{3}&=&\rho\frac{\omega_{41}~\omega_{12}^{f}~\omega_{23}Q}{K}\\
P_{4}&=&\rho\frac{\omega_{12}^{f}~\omega_{23}~\omega_{34}Q}{K}
\end{eqnarray}
Where;
\begin{eqnarray}
K&=&\omega_{23}\omega_{34}\omega_{41}+\omega_{34}\omega_{41}\omega_{21}^{b}Q +\omega_{34}\omega_{41}\omega_{12}^{f}Q \nonumber \\ 
&+&\omega_{41}\omega_{12}^{f}\omega_{23}Q+\omega_{12}^{f}\omega_{23}\omega_{34}Q
\end{eqnarray}
The corresponding flux $J$ is given by
\begin{eqnarray}
J&=&~Q\biggl[~\left(\omega^f_{12}-\omega^b_{11} \right)P_{1}-\omega^b_{21}P_{2}\biggr]\nonumber\\
&=&\dfrac{\rho~Q }{K}~\omega_{34}~\omega_{41}\biggl[~\omega_{12}^f
~\omega_{23} - \omega_{11}^b~ \omega_{23} - \omega_{11}^b~\omega_{21}^b~Q \biggr]\nonumber \\
\label{eq-pj}
\end{eqnarray}
Hence, the average velocity $V$ of a single RNAP is given by
\begin{eqnarray}
V=J/\rho =\dfrac{~Q }{K}~\omega_{34}~\omega_{41}\biggl[~\omega_{12}^f
~\omega_{23} - \omega_{11}^b~ \omega_{23} - \omega_{11}^b~\omega_{21}^b~Q \biggr]\nonumber \\
\label{eq-pv}
\end{eqnarray}
The average velocity $V$ depends on the number density $\rho$ through 
the $\rho$-dependence of $Q$. 

In the following subsections we'll use the formula (\ref{eq-pv}) in 
the regime of sufficiently low coverage density $\rho_{cov}$ to 
predict the dependence of $V$ of single RNAP motors on external force 
and torque. Such dependences of $V$ on external force and torque can 
be probed by carrying out single-molecule experiments {\it in-vitro}.
In order to predict the total rate of RNA synthesis at coverage 
densities where steric interactions between the RNAPs is not negligibly 
small, we'll use the expression (\ref{eq-pj}). Thus, our model may be 
regarded as a ``unified'' description of transcription in the sense 
that it can account for properties of single RNAP motors as well as 
their collective behaviour.

%%%%%%%%%%%%%%%%%%%%%%%%%%%%%%%%%%%%%%%%%%%%%%%%%%%%%%%%%%%%%%%%%%
For numerical calculations, we use the rate constants extracted by 
Wang et al.\cite{wangrnap} from the empirical data. In the absence 
of external force and torque, the typical numerical values of these 
rate constants are as follows:
%%%%%%%%%%%%%%%%%%%%%%%%%%%%%%%%%%%%%%%%%%%%%%%%%%%%%%%%%%%%%%%%%%%%%%%
\begin{eqnarray}
\omega^{b}_{11} &=& 9.4 ~s^{-1} \nonumber \\
\omega^{f}_{12}&=&\omega^{f0}_{12} \cdot [NTP], ~{\rm with} ~\omega^{f0}_{12} = 10^{6} ~M^{-1} \cdot s^{-1} \nonumber \\ 
\omega^{b}_{21}&=&0.21 s^{-1} \nonumber \\
\omega_{23} &=&100.0~ s^{-1}  \nonumber \\
\omega_{34}&=&1000.0 ~s^{-1} \nonumber\\
\omega_{41} &=& 31.4 ~s^{-1} \nonumber \\
\end{eqnarray}
So far as the two new rate constants $\omega_{23}$ and $\omega_{41}$ 
are concerned, we investigate their effects on the rate of transcription 
by varying their magnitudes over a wide range.

%%%%%%%%%%%%%%%%%%%%%%%%%%%%%%%%%%%%%%%%%%%%%%%%%%%%%%%%%
\subsection{Effects of externally applied force and torque}
%%%%%%%%%%%%%%%%%%%%%%%%%%%%%%%%%%%%%%%%%%%%%%%%%%%%%%%%%

In this section we examine the effects of external force $F$ and torque 
$\tau$ on the rate of RNA polymerization. From the force-velocity relation, 
we extract the stall force $F_{s}$, the load force which just stalls the 
a single RNAP motor. In order to extract these single RNAP properties 
from our model, we first use the formula (\ref{eq-pv}) in the regime of 
extremely low coverage density of the RNAP motors.

%%%%%%%%%%%%%%%%%%%%%%%%%%%%%%%%%%%%%%%%%%%%%%%%%%%%%%%%%%%%%%%%%%%%%%%
\begin{figure}[h]
\begin{center}
\includegraphics[width=0.95\columnwidth]{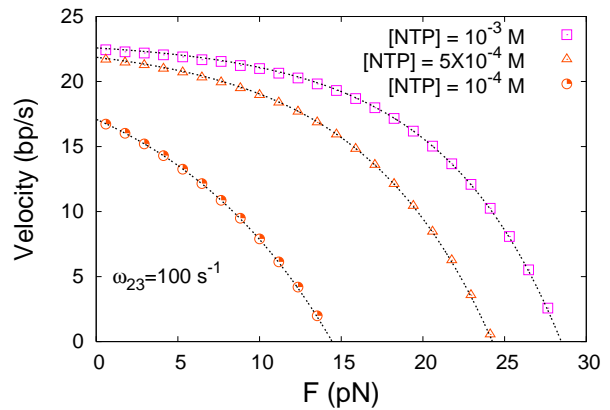}\\[0.5cm]
\end{center}
\caption{The average speed of a single RNAP motor is plotted against the 
load force (a) for three different values of the NTP concentration. 
The lines have been obtained from the analytical expression (\ref{eq-pv}) 
in extremely low number density of the RNAP motors. The discrete data  
points have been obtained by computer simulations of the model under 
identical conditions.
}
\label{fig-vvsf}
\end{figure}
%%%%%%%%%%%%%%%%%%%%%%%%%%%%%%%%%%%%%%%%%%%%%%%%%%%%%%%%%%%%%%%%%%%%%

%%%%%%%%%%%%%%%%%%%%%%%%%%%%%%%%%%%%%%%%%%%%%%%%%%%%%%%%%%%%%%%%%%%%%%%
\begin{figure}[h]
\begin{center}
\includegraphics[width=0.95\columnwidth]{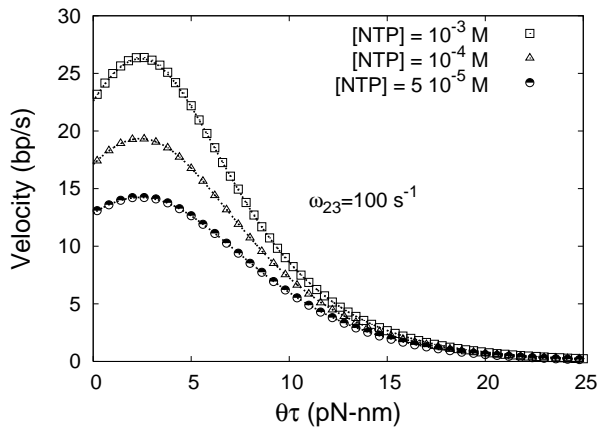}\\[0.5cm] 
(a)
\includegraphics[width=0.95\columnwidth]{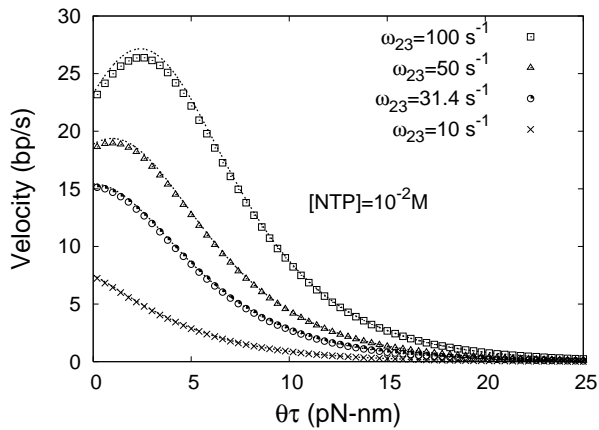}\\[0.5cm] 
(b)
\end{center}
\caption{The average speed of a single RNAP motor is plotted against the 
torque (a) for three different values of the NTP concentration for fixed 
$\omega_{23} = 100$s$^{-1}$, (b) for four different values of the rate 
constant $\omega_{23}$, keeping NTP concentration fixed at $10^{-2}M$. 
The lines have been obtained from the analytical expression (\ref{eq-pv}) 
at extremely low coverage density of the RNAP motors. The discrete data  
points have been obtained by computer simulations of the same model under 
identical conditions.
}
\label{fig-vvst}
\end{figure}
%%%%%%%%%%%%%%%%%%%%%%%%%%%%%%%%%%%%%%%%%%%%%%%%%%%%%%%%%%%%%%%%%%%%%

%%%%%%%%%%%%%%%%%%%%%%%%%%%%%%%%%%%%%%%%%%%%%%%%%%%%%%%%%
\subsubsection{Effects of external force}
%%%%%%%%%%%%%%%%%%%%%%%%%%%%%%%%%%%%%%%%%%%%%%%%%%%%%%%%%

Following the standard practice in the master equation approach to 
molecular motors \cite{fisher} for capturing the effects of external 
force on the mechano-chemistry \cite{busta04}, we multiply the 
``mechanical'' steps of the cycle by an appropriate Boltzmann-like 
factor. More precisely,
we assume that the load force $F$ significantly affects only those steps 
of the mechano-chemical cycle of an RNAP which involve its mechanical 
movement, i.e., forward or backward movement in real space. Therefore, 
we assume that force-dependence of these rate constants are given by 
\begin{eqnarray}
\omega_{12}^{f}(F) &=&\omega_{12}^{f}~ exp(- F ~\delta/k_{B}T) \nonumber \\ 
\omega_{21}^{b}(F)&=&\omega_{21}^{b}~ exp(F ~\delta/k_{B}T) \nonumber \\
\omega_{11}^{b}(F)&=&\omega_{11}^{b}~ exp(F ~\delta/k_{B}T) 
\label{eq-fdeprates}
\end{eqnarray}
where $\omega_{12}^{f}, \omega_{21}^{b}$ and $\omega_{11}^{b}$ are 
the corresponding values of the rate constants in the absence of the 
load force. In equation (\ref{eq-fdeprates}) the symbol $\delta = 0.34$ nm 
is the typical length of a single nucleotide. None of the rate constants 
other than the three listed in (\ref{eq-fdeprates}) are affected by the 
load force $F$. 

The force-velocity relation of individual RNAP motors in our model is 
shown in fig.\ref{fig-vvsf} for three different concentrations of the 
NTPs. The average velocity decreases monotonically with increasing 
load force; the convex shape of the load-velocity curves are very similar 
to those reported by Wang et al. \cite{wangrnap} in their single RNAP 
model. In this model, the higher is the NTP concentration the larger is 
the stall force $F_s$. Moreover, for a given load force $F < F_s$, the 
average velocity of the RNAP is larger at higher NTP concentration.

%%%%%%%%%%%%%%%%%%%%%%%%%%%%%%%%%%%%%%%%%%%%%%%%%%%%%%%%%
\subsubsection{Effects of external torque}
%%%%%%%%%%%%%%%%%%%%%%%%%%%%%%%%%%%%%%%%%%%%%%%%%%%%%%%%%

Now we consider the effects of an externally imposed torque which assists 
opening, but opposes closing of the {\it palm}-like shape of an isolated 
RNAP.  These competing effects of the same torque on the two steps of each 
mechano-chemical cycle of individual RNAP motors has nontrivial effects 
on the rate of RNA synthesis. The above mentioned effects of the torque 
$\tau$ on opening and closing of RNAP are captured by the following choice 
of the corresponding rate constants 
\begin{eqnarray}
\omega_{23}(\tau)&=&\omega_{23}~ exp(-\tau~\theta/k_{B}T) \nonumber \\
\omega_{41}(\tau)&=&\omega_{41}~exp(\tau~\theta/k_{B}T) 
\end{eqnarray}
where $\omega_{23}$ and $\omega_{41}$ on the right hand sides of the 
equations are the rate constants in the absence of external torque.

The average velocity of the RNAP motors are plotted against the torque 
in fig.\ref{fig-vvst} for different sets of values of the model parameters. 
The most dramatic effect of the torque is the {\it nonmonotonic} 
variation of $V$ with $\tau$ in several experimentally accessible 
regimes of parameter values. Increasing torque increases the opening 
rate and decreases the closing rate. As long as the opening rate is 
still lower (and, hence, rate limiting), the velocity increases with 
increasing torque. A peak appears where closing becomes the rate 
limiting step. In order to demonstrate the effect of this crossover 
from a regime dominated by opening to that dominated closing of RNAP, 
we have plotted one curve in fig.\ref{fig-vvst}(b) corresponding to  
$\omega_{23} = \omega_{41}$ for which the maximum occurs at $\tau = 0$.

%%%%%%%%%%%%%%%%%%%%%%%%%%%%%%%%%%%%%%%%%%%%%%%%%%%%%%%%%
\subsection{Effects of steric interactions: flux-density relation}
%%%%%%%%%%%%%%%%%%%%%%%%%%%%%%%%%%%%%%%%%%%%%%%%%%%%%%%%%
%%%%%%%%%%%%%%%%%%%%%%%%%%%%%%%%%%%%%%%%%%%%%%%%%%%%%%%%%%%%%%%%%%%%%%%%%%%%%%%%%%

%%%%%%%%%%%%%%%%%%%%%%%%%%%%%%%%%%%%%%%%%%%%%%%%%%%%%%%%%%%%%%%%%%%%%%%%%%%%%%
\begin{figure}
\includegraphics[width=0.75\columnwidth]{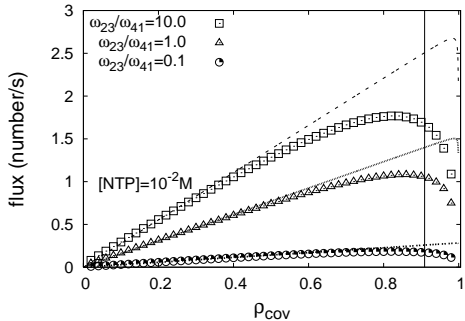}
\caption{The steady-state flux of the RNAPs, under periodic boundary 
conditions, plotted as a function of the coverage density $\rho_{cov}$ 
for three different values of the ratio $\omega_{23}/\omega_{41}$. The 
vertical line represents the position of our theoretical estimate 
(\ref{eq-max}) of the maximum flux for fixed value of transition rates.}
\label{fig-pbc}
\end{figure}
%%%%%%%%%%%%%%%%%%%%%%%%%%%%%%%%%%%%%%%%%%%%%%%%%%%%%%%%%%%%%%%%%%%%%%%%%%%%%%

We have plotted mean-field estimate (\ref{eq-pj}) of flux $J$ against 
the coverage density $\rho_{cov}$ in fig.\ref{fig-pbc} for three  
different valus of the ratio $\omega_{23}/\omega_{41}$ at 
fixed NTP concentration of $10^{-2} M$. In the traffic science 
literature \cite{css} such flux-density relations are usually referred 
to as the fundamental diagram. The qualitative features of the 
fundamental diagrams in fig.\ref{fig-pbc} are similar to those 
observed earlier in the two state model of RNAP traffic \cite{ttdc}.  
The most notable feature of these fundamental diagrams is their 
{\it asymmetric shape}; this is in sharp contrast to the symmetry of  
fundamental diagram of a TASEP about $\rho = 1/2$. The physical reason 
for the asymmetric shape of the fundamental diagram in fig.\ref{fig-pbc}  
is as follows:\
to get maximum current, system should have maximum number of particle-hole
pairs. For $r=1$, maximum number of particle hole pairs occur when 
$1-\rho = \rho$, i.e., $\rho=\rho^{m}=0.5$. For $r\geq 2$, the 
magnitude of $\rho^{m}$ can be found from the corresponding condition 
$1-\rho^{m} ~r= \rho^{m}$, i.e.,  $\rho^{m} = 1/(r+1)$, which corresponds 
to the coverage density 
\begin{equation}
\rho_{cov}^{m} =r/(r+1). 
\label{eq-max} 
\end{equation}

%%%%%%%%%%%%%%%%%%%%%%%%%%%%%%%%%%%%%%%%%%%%%%%%%%%%%%%%%%%%%%%%%%%%%%%%%%%%%%%%%

%%%%%%%%%%%%%%%%%%%%%%%%%%%%%%%%%%%%%%%%%%%%%%%%%%%%%%%%%%%%%%%%%%%%%%%%%%%
\section{RNAP TRAFFIC MODEL UNDER OPEN BOUNDARY CONDITION}
\label{sec-obc}
%%%%%%%%%%%%%%%%%%%%%%%%%%%%%%%%%%%%%%%%%%%%%%%%%%%%%%%%%%%%%%%%%%%%%%%%%%%

For modeling transcription, the open boundary condition is more realistic 
than the periodic boundary conditions. Under this conditon a RNAP can 
attach to start site(labled as $n=1$) with rate $\omega_{\alpha}$ provided  
none of the first $r$ sites on the lattice is covered by any other RNAP. 
Similarly, after reaching the site $n = r$, an RNAP leaves the DNA track 
with rate $\omega_{\beta}$. We calculate the conditional probability $Q$ 
same way as we have done for the 2-state model of RNAP \cite{ttdc}. 
In this case, under mean-field approximation, the master equations for the 
probabilities $P_{\mu}(n,t)$ are given by

%%%%%%%%%%%%%%%%%%%%%%%%%%%%%%%%%%%%%%%%%%%%%%%%%%%%%%%%%%%%%%%%%%%%%%%%%%%%%%%%%%%%
\begin{widetext}
\begin{eqnarray}
\frac{dP_{1}(n,t)}{dt}&=&  ~\omega_{\alpha} ~\biggl(1-\sum_{s=1}^{r}~P(s)\biggr)~\delta(n-1)+\biggl[~\omega^{b}_{11} ~P_{1}(n+1,t) + ~\omega^{b}_{21} ~P_{2}(n+1,t)\biggr]\biggl( 1-\textsl{H}(n-L)\biggr) ~Q^{-}(n+1)\nonumber\\
&+&~\omega_{41} ~P_{4}(n,t)- ~\omega^{b}_{11} ~P_{1}(n,t)~\textsl{H}(n-2) ~Q^{-}(n)\nonumber\\
&-&~\omega^{f}_{12}~P_{1}(n,t)\biggl( 1-\textsl{H}(n-L)\biggr)~Q^{+}(n)-\omega_{\beta}~P_{1}(n,t)~\delta(n-L)\label{eq-obc1}\\
\frac{dP_{2}(n,t)}{dt} &=&~\omega^{f}_{12} ~P_{1}(n-1,t)~\textsl{H}(n-2)~Q^{+}(n-1)
-~\omega_{23}~P_{2}(n,t)\nonumber\\
&-&~\omega^{b}_{21}~P_{2}(n,t)~\textsl{H}(n-2)~Q^{-}(n)\label{eq-obc2}\\
\frac{dP_{3}(n,t)}{dt} &=&~\omega_{23} ~P_{2}(n,t)-\omega_{34} ~P_{3}(n,t)\label{eq-obc3}\\
\frac{dP_{4}(n,t)}{dt}&=&\omega_{34} ~P_{3}(n,t)-\omega_{41} ~P_{1}(n,t)\label{eq-obc4}
\end{eqnarray}
\end{widetext}
Where $H(n)$ is the Heaviside step function defined by 
\begin{eqnarray}
\textsl{H}(n)  = \left\{ \begin{array}{ll}
         1 & \mbox{if $n \geq 0$}\\
         0 & \mbox{if $n < 0$}.\end{array} \right. \nonumber\\
\end{eqnarray}
and
\begin{eqnarray}
Q^{-}(n) = \left\{ \begin{array}{ll}
         1 & \mbox{if $n \leq r$}\\
         \biggl(\dfrac{1-\sum_{s=1}^{r} P(n-s)}{1-\sum_{s=1}^{r} P(n-s)+P(n-r)}\biggr) & \mbox{if $n > r$}.\end{array} \right.\nonumber\\
\end{eqnarray}
\begin{eqnarray}
Q^{+}(n) = \left\{ \begin{array}{ll}
         1 & \mbox{if $n > L-r$}\\
         \biggl(\dfrac{1-\sum_{s=1}^{r} P(n+s)}{1-\sum_{s=1}^{r} P(n+s)+P(n+r)}\biggr) & \mbox{if $n \leq L-r$}.\end{array} \right.\nonumber\\
\end{eqnarray}

%%%%%%%%%%%%%%%%%%%%%%%%%%%%%%%%%%%%%%%%%%%%%%%%%%%%%%%%%%%%%%%%%%%%%%%
\begin{figure}[h]
\begin{center}
\includegraphics[width=0.95\columnwidth]{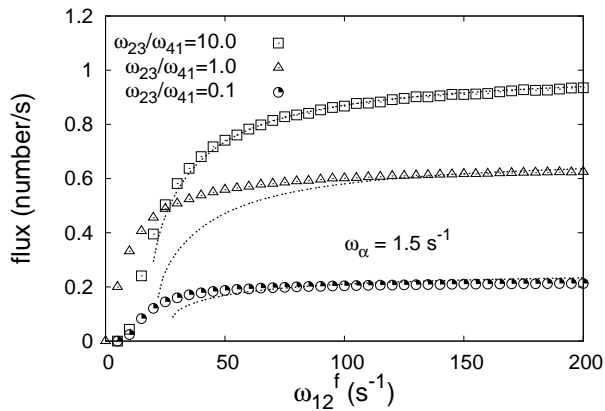}\\[0.5cm]
\end{center}
\caption{The steady state current of RNAPs under open boundary condition 
for three different values of ratio $\omega_{23}/\omega_{41}$ .
}
\label{fig-fdobc}
\end{figure}
%%%%%%%%%%%%%%%%%%%%%%%%%%%%%%%%%%%%%%%%%%%%%%%%%%%%%%%%%%%%%%%%%%%%%

We have numerically solved the mean-field equations 
(\ref{eq-obc1})-(\ref{eq-obc4}) in the steady-state to compute the 
corresponding flux of the RNAP motors. These mean-field theoretic 
estimates of flux are compared with the corresponding data obtained 
from direct computer simulations of the model. These results are 
plotted as functions of the rate constants $\omega_{\alpha}$ and 
$\omega^{f}_{21}$, respectively, in figs.\ref{fig-fdobc} (a) and (b). 
The flux rises with increasing $\omega_{12}^{f}$, but the rate of 
rise decreases with increasing $\omega_{12}^{f}$. Eventually, the 
flux saturates because $\omega_{12}^{f}$ is no longer rate-limiting. 
This trend of variation of the total rate of RNA synthesis with the 
concentration of NTP is similar to that observed earlier in our 
2-state model of RNAP traffic \cite{ttdc}. Thus, a 2-state model 
of RNAP would be adequate to capture the dependence of the rate of 
RNA synthesis on the NTP concentration.

%%%%%%%%%%%%%%%%%%%%%%%%%%%%%%%%%%%%%%%%%%%%%%%%%%%%%%%%%%%%%%%%%%
\section{Summary and conclusions}
%%%%%%%%%%%%%%%%%%%%%%%%%%%%%%%%%%%%%%%%%%%%%%%%%%%%%%%%%%%%%%%%%%

In this paper we have extended our earlier 2-state model of RNAP traffic 
to a 4-state model so as to capture some important cyclic shape changes  
in the mechano-chemical cycle of each RNAP. The new model predicts the 
effects of these shape changes on the rate of RNA synthesis. Moreover, 
we have used the same model in the extremely low density limit to extract 
the force-velocity relation. Finally, we have demonstrated a novel 
nonmonotonic variation of the average speed of the RNAP motors (and, 
hence, a nonmonotonic variation of the rate of synthesis of RNA) with the 
increase of an externally imposed torque on the individual motors. 
Nonmonotonic variation of average speed with external load force 
(not external torque) has been observed earlier \cite{goel1,goel2,goel3} 
in the case of several DNA polymerases. But, the physical origin of 
this nonmonotonicity, as interpreted in ref.\cite{goel1,goel2,goel3}, 
is quite different from that responsible for the nonmonotonic variation 
of average speed with external torque. Moreover, the present version 
of our model does not explicitly capture the effect of possible twisting 
of the DNA by the polymerase.
In principle, it should be possible to test the new predictions of our 
model by carrying out laboratory experiments {\it in-vitro}. 

To our knowledge, it may not be possible, at present, to apply a torque 
that would directly oppose (or assist) the opening (or closing) of a 
RNAP. Therefore, we now suggest a possible alternative technique. The 
nonmonotonic variation of the velocity of the RNAP with external torque 
will result also in the case where the torque opposes, rather than 
assisting, the opening of the individual RNAP motors (i.e., opposite to 
the direction of the torque shown in fig.\ref{fig-architec}). Suppose, 
the finger-like domain of each RNAP is genetically cross-linked with the  
palm domain by a protein whose entropic elastic constant is significant. 
When such an RNAP tends to open up, the cross-linking protein opposes 
the opening. By repeating the experiment with different cross-linkers, 
whose entropic elastic constants are different, one can study the variation 
of the average velocity of the RNAP motors with the applied torque.

In spite of some crucial differences, most of the polynucleotide 
polymerases seem to share a common ``cupped right hand'' architecture  
\cite{cramer02b}. Therefore, it should be possible to make minor 
modifications in our 4-state model of RNAP so as to develop similar 
models of other polynucleotide polymerases.\\

\noindent{\bf Acknowledgements}: This work is supported by a research 
grant from CSIR (India). 

%%%%%%%%%%%%%%%%%%%%%%%%%%%%%%%%%%%%%%%%%%%%%%%%%%%%%%%%%%%%%%%%%%%%%%%%

%%%%%%%%%%%%%%%%%%%%%%%%%%%%%%%%%%%%%%%%%%%%%%%%%%%%%%%%%%%%%%%%%%%%%%%%

\end{document}